\documentclass{article}

\usepackage{arxiv}

\usepackage[utf8]{inputenc} 
\usepackage[T1]{fontenc}    
\usepackage{hyperref}       
\usepackage{url}            
\usepackage{booktabs}       
\usepackage{amsfonts}       
\usepackage{nicefrac}       
\usepackage{microtype}      
\usepackage{lipsum}
\usepackage{graphicx}
\graphicspath{ {./images/} }

\title{AxonCallosumEM Dataset: Axon Semantic Segmentation of Whole Corpus Callosum cross section from EM Images}

\author{
  Ao Cheng \\
  School of Electronic and Information Engineering\\
  Soochow University\\
  Suzhou 215009, China\\
  Jiangsu Key Laboratory of Medical Optics\\
  Suzhou Institute of Biomedical Engineering and Technology\\
  Chinese Academy of Sciences\\
  Suzhou 215163, China\\
  \texttt{aoc.7@outlook.com} \\
  \And
  Guoqiang Zhao\\
  School of Biomedical Engineering (Suzhou),\\
  Division of Life Sciences and Medicine,\\
  University of Science and Technology of China\\
  Suzhou, China\\
  \texttt{zhaoguoqiang@mail.ustc.edu.cn} \\
   \And
  Lirong Wang\\
  School of Electronic and Information Engineering\\
  Soochow University\\
  Suzhou 215009, China\\
  \texttt{wanglirong@suda.edu.cn} \\
  \And
  Ruobing Zhang \\
  Jiangsu Key Laboratory of Medical Optics\\
  Suzhou Institute of Biomedical Engineering and Technology\\
  Chinese Academy of Sciences\\
  Suzhou 215163, China\\
  \texttt{zhangrb@sibet.ac.cn} \\
}

\begin{document}
\maketitle
\begin{abstract}
The electron microscope (EM) remains the predominant technique for elucidating intricate details of the animal nervous system at the nanometer scale. However, accurately reconstructing the complex morphology of axons and myelin sheaths poses a significant challenge. Furthermore, the absence of publicly available, large-scale EM datasets encompassing complete cross sections of the corpus callosum, with dense ground truth segmentation for axons and myelin sheaths, hinders the advancement and evaluation of holistic corpus callosum reconstructions. To surmount these obstacles, we introduce the AxonCallosumEM dataset, comprising a 1.83×5.76mm EM image captured from the corpus callosum of the Rett Syndrome (RTT) mouse model, which entail extensive axon bundles. We meticulously proofread over 600,000 patches at a resolution of 1024×1024, thus providing a comprehensive ground truth for myelinated axons and myelin sheaths. Additionally, we extensively annotated three distinct regions within the dataset for the purposes of training, testing, and validation. Utilizing this dataset, we develop a fine-tuning methodology that adapts Segment Anything Model (SAM) to EM images segmentation tasks, called EM-SAM, enabling outperforms other state-of-the-art methods. Furthermore, we present the evaluation results of EM-SAM as a baseline.
\end{abstract}

\section{Introduction}
The intricate network of nerves within the brain forms a complex web of interactions, enabling the remarkable cognitive abilities exhibited by higher organisms. Understanding the structural organization of these nerves and their interplay is of paramount importance in deciphering the functioning of the brain. With recent advancements in electron microscopy (EM) technology, it has become feasible to acquire neuronal structures of brain \cite{eberle2015em}. Utilizing high-throughput and accurate segmentation methods \cite{ronneberger2015unet,lee2017superhuman}, several gray matter EM datasets have been reconstructed and published \cite{lucchi2011supervoxel,wei2020mitoem,wei2021axonem}. However, only a few unreleased EM dataset \cite{abdollahzadeh2021deepacson} focusing on the white matter, corpus callosum, which is a prominent structure interconnecting the two cerebral hemispheres, enables the exchange of information and coordination of various cognitive processes between the brain's left and right sides and contains vast quantities of axon bundles with complex morphology. 

\begin{figure}
  \centering
  \includegraphics[width = 1\textwidth]{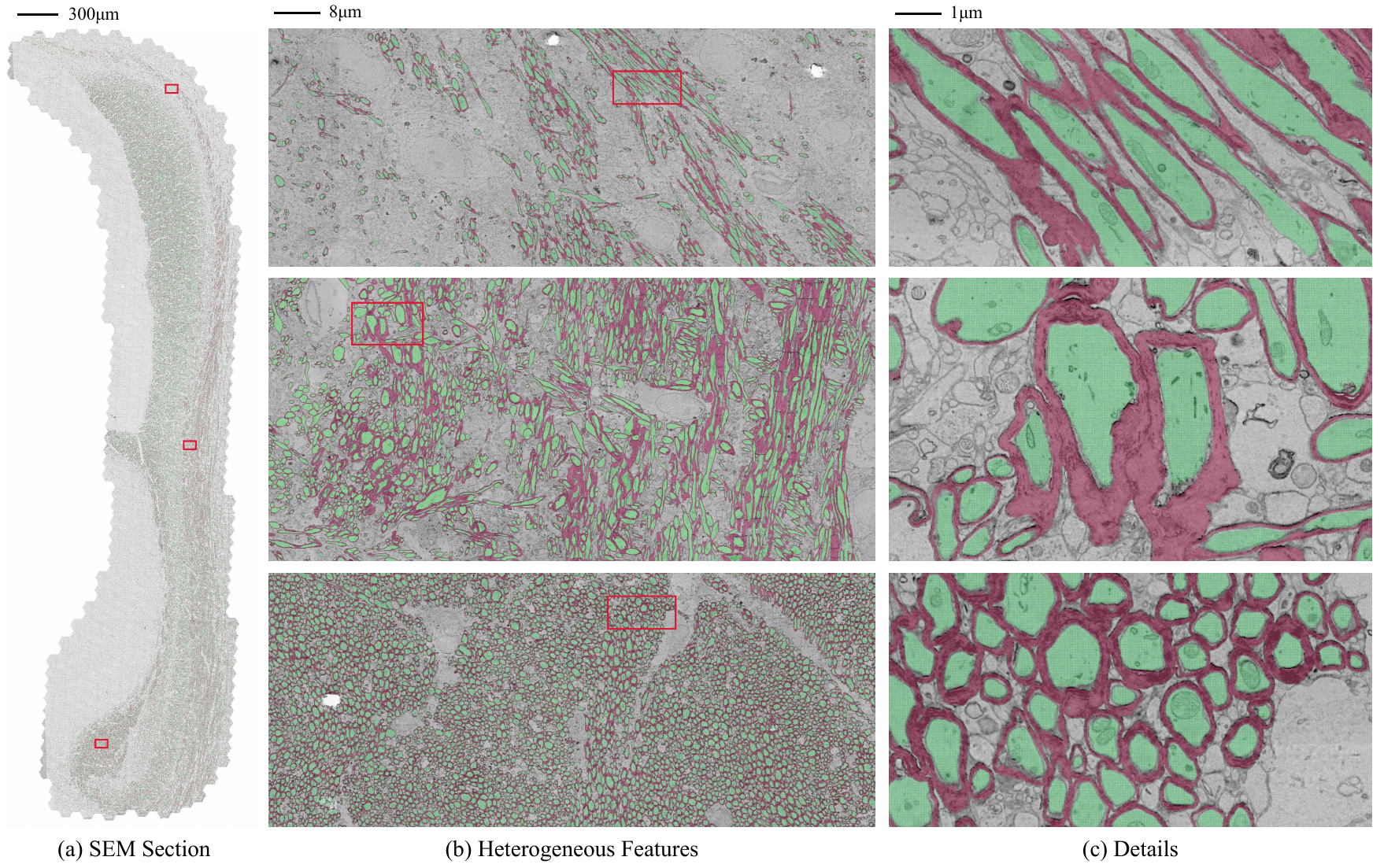}
  \caption{Myelinated axons and their corresponding myelin sheaths in the AxonCallosumEM dataset. (a) Reconstruction outcome of our dataset, showcasing three distinct regions: the genu (top), the body (middle), and the splenium (bottom) of the corpus callosum, selected for further illustration. (b) The distribution of myelinated axons exhibits heterogeneous characteristics. In the genu, axons are sparsely distributed. In the body, axons display diverse morphologies. In the splenium, axons are densely distributed. (c) The genu, body, and splenium of the corpus callosum exhibit distinct features based on their detailed analysis, including long-range axons, thicker myelin sheaths, and dense axon distributions, respectively.}
  \label{fig1}
\end{figure}

To the best of our knowledge, no existing EM reconstruction from any whole corpus callosum cross section provides densely and fully proofread ground truth for myelinated axons and myelin sheaths. Moreover, current state-of-the-art methods from other public EM dataset cannot overcome the challenge of corpus callosum reconstruction task. As shown in Fig. \ref{fig1}, the morphology of axons significantly varies across three regions of the corpus callosum, making state-of-the-art neuron segmentation methods based on convolutional neural networks (CNN) inadequate. In Fig. \ref{fig2} (a), we demonstrate the challenge posed by scale differences and the dense distribution of myelinated axons, which require segmentation methods with large perceptual fields to accurately categorize pixels as inner cell or background. Additionally, Fig. \ref{fig2} (b) highlights the need for segmentation methods to fully understand the semantic information of myelinated axons and their myelin sheaths due to the presence of special cases such as demyelination and the Node of Ranvier. Finally, Fig. \ref{fig2} (c) showcases the long-range axons and complex morphology present in the corpus callosum. Regions with long-range axons often lack distinct boundary features, necessitating the development of more robust methods to differentiate these axons from the common circular axons. Therefore, we need a large-scale corpus callosum dataset to evaluate current methods and foster new researches to address the these challenges. 

To aim this, we have created a large-scale benchmark for a 2D axon semantic segmentation dataset called AxonCallosumEM. Our dataset is the Rett Syndrome (RTT) mouse model \cite{vashi2019treating}, comprising an extensive collection of over 600,000 patches. Each patch is rendered at a resolution of $1024\times1024$ with a pixel size of 4 nm. Moreover, we provide a comprehensive ground truth for myelinated axons and myelin sheaths. Our dataset demonstrates the reconstruction of the entire corpus callosum cross section and showcases various morphologies and distributions of myelinated axons and myelin sheath, enabling further comparative studies across different mammalian species and mouse models. Furthermore, we have fine-tuned the state-of-the-art nature images segmentation model, Segment Anything Model (SAM \cite{kirillov2023segment}), to adapt specifically to EM segmentation tasks, thereby transforming it into EM-SAM. This adaptation effectively overcomes the majority of challenges encountered in AxonCallosumEM, while also presenting opportunities for exploring the adaptability of models originally designed for nature images to the biomedical applications.

\subsection{Related work}
For the segmentation tasks, conventional methods such as morphology-based processing \cite{cuisenaire1999automatic,romero2000automatic,more2011semi}, region growing and labeling \cite{zhao2010automatic}, watershed \cite{beucher1992watershed,wang2012segmentation}, and contours-based analysis \cite{begin2014automated,zaimi2016axonseg} are proposed. Recently, deep-learning based methods (e.g UNet\cite{ronneberger2015unet}, FPN\cite{lin2017fpn}) have been proven capable of EM segmentation tasks and outperform the conventional methods. Moreover, predicting affinity map \cite{lee2017superhuman} and additional targets \cite{wei2020mitoem,lin2021nucmm} are proposed for further improvement. Such approaches are used to process 2D \cite{zaimi2018axondeepseg} and 3D \cite{motta2019dense,abdollahzadeh2021deepacson,wei2021axonem} EM datasets. However, as the limitation of perceptive field from convolution kernel, it always fails on segmenting large-scale cellular structures.

More recently, vision transformer (Vit) \cite{Vaswani2017transformer,Dosovitskiy2021google_transformer_experiments}, which is first used in classification tasks and are adopted from sequence-to-sequence modeling in natural language processing, are proposed to solve such issues utilizing its long-range dependency. In the field of medical image segmentation, Vit-based \cite{hatamizadeh2022unetr,hatamizadeh2022swinunetr} network utilized the transformer blocks as strong encoder and outperform CNN-based methods. In order to maximize the capability of feature extraction of vit-based encoder, Tang \emph{et al.} \cite{tang2022self3dswin} and Cheng \emph{et al.} \cite{cheng2023learning} proposed the masked image modeling (MIM) pre-training and fine-tuning paradigm on Swin \cite{liu2021swin} and Vit \cite{Dosovitskiy2021google_transformer_experiments} respectively. The MIM method (e.g DAE \cite{vincent2008dae,vincent2010dae}, MAE\cite{he2021mae}, BeiT\cite{bao2021beit,peng2022beit}) is a self-supervised learning method that learns representations from the image itself. In the field of CV, the pre-training methods \cite{zhou2021ibot,Wei_2022_maskedfeat,xie2022simmim} continue to develop and several large models \cite{caron2021dino,oquab2023dinov2,kirillov2023segment} have proven the domination on downstream tasks.

\section{AxonCallosumEM dataset}
The proposed AxonCallosumEM dataset encompasses the entire cross section of the mouse corpus callosum. We conducted dense annotations of myelinated axons and myelin sheaths in order to evaluate semantic segmentation methods and facilitate biomedical analysis.

\begin{figure}
  \centering
  \includegraphics[width = 1\textwidth]{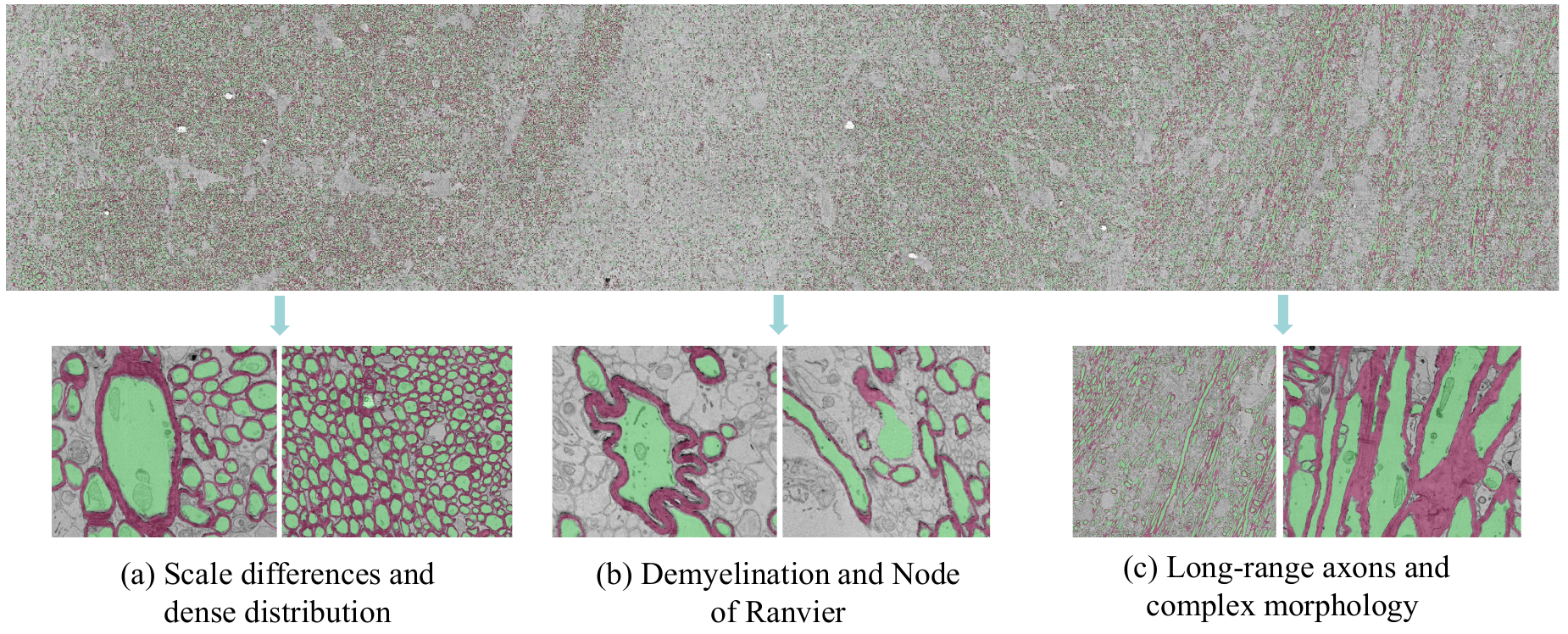}
  \caption{Demonstration of challenges in our AxonCallosumEM dataset, focusing on the splenium of the corpus callosum: (a) Scale differneces and dense distribution. (b) Demylination and Node of Ranvier. (c) Long-range axons and complex morphology. These complex cases are widespread, but remain unaddressed in existing datasets.}
  \label{fig2}
\end{figure}

\subsection{Dataset discription}
To construct the dataset, we utilized the Rett Syndrome (RTT) mouse model, specifically the Mecp2-mutant mice \cite{vashi2019treating}. The EM images were acquired at a resolution of 4nm per pixel, encompassing the entire cross section of the corpus callosum, including the genu, the body, and the splenium of the corpus callosum \cite{johnson2023merged}. The dataset was rendered with a patch size of $1024\times1024$ at 4nm per pixel resolution. Overall, the dataset comprises 448 and 1408 patches along the x and y directions, respectively, resulting in a total of over 600,000 patches. 

\subsection{Dataset annotation}
For reconstruction purposes, we adopted a semi-automatic approach. Initially, we manually annotated myelinated axons and myelin sheaths within a region of $224\times160$ patches, with an 8nm per pixel resolution. Considering the complex morphology of axons, the annotation process commenced across the entire x-axis, as the features exhibited significant variations (see Fig. \ref{fig1}) along the y-axis. Subsequently, we fine-tuned the image encoder of the vit-based model \cite{Vaswani2017transformer,Dosovitskiy2021google_transformer_experiments} using the public released SAM model \cite{kirillov2023segment} (SAM-Base) with the annotated labels. This allowed us to predict the remaining region along the y-axis. The proofreader then inspected and manually corrected errors in the VAST \cite{berger2018vast}. Following this pipeline, we iteratively accumulated ground truth semantic segmentation by gradually expanding the annotated patches along the y-axis to reconstruct the entire cross section of the corpus callosum. 

\subsection{Dataset analysis}
\begin{figure}
  \centering
  \includegraphics[width = 1\textwidth]{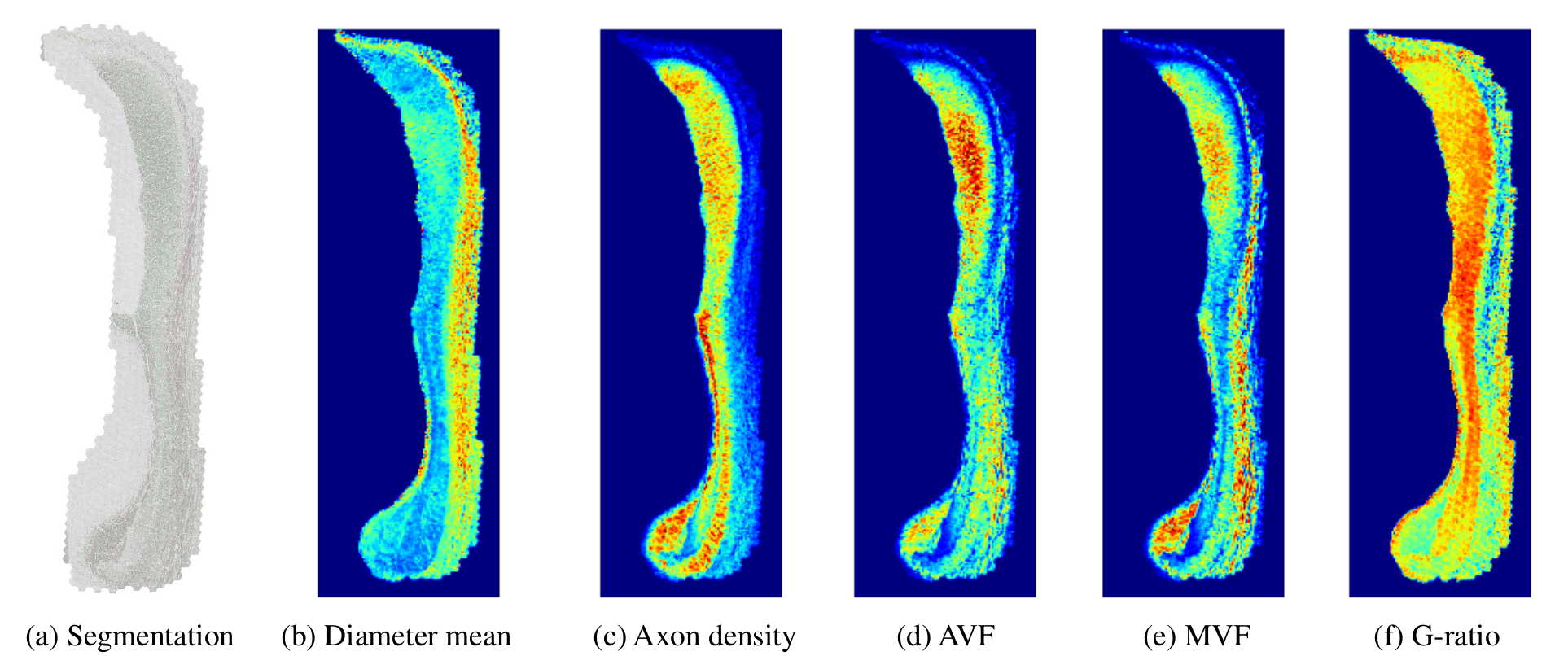}
  \caption{Distribution maps depicting various metrics within the cross-section of the entire corpus callosum, including (b) mean and standard deviation of axon diameter, (c) axon density, (d) axon volume fraction, (e) myelin volume fraction, and (f) g-ratio. The aggregate measurements of the white matter were obtained by downsampling the labeled data to a resolution of 16 nm per pixel, with a patch size of $1024\times1024$. To improve visual contrast, a max-min normalization technique was applied to all the obtained outcomes.}
  \label{fig3}
\end{figure}

The segmented mouse corpus callosum is shown in Fig. \ref{fig3} (a). For biomedical analysis, we downsampled the labels to a resolution of 16nm per pixel. Additionally, a binary mask was applied to exclusively retain the region of the corpus callosum for subsequent computations, based on a patch size of $1024\times1024$. The following aggregate metrics \cite{zaimi2018axondeepseg} were computed:
\begin{enumerate}
\item{\textbf{Axon diameter mean and standard deviation}. Arithmetic mean and standard deviation of the distribution of equivalent axon diameters (computed for each axon object as $\sqrt{(4\times Area/\pi)}$).}
\item{\textbf{Axon density}. Number of axons per patches.}
\item{\textbf{Axon volume fraction (AVF)}. The ratio between area of axons and total area of the region.}
\item{\textbf{Myelin volume fraction (MVF)}. The ratio between area of myelin and total area of the region.}
\item{\textbf{G-ratio}. The ratio between axon diameter and myelinated fiber $(axon + myelin)$ diameter, which can be estimated with the following: $1/\sqrt{1+(MVF/AVF)}$.}
\end{enumerate}

The results are shown in Fig. \ref{fig3} (b) to (f). To enhance contrast, we applied max-min normalization to all the results. It was observed that the average g-ratio was approximately 0.7. Note that the distribution map of g-ratio of the RTT mouse model exhibited different results with known anatomy of other mouse models \cite{west2018experimental,arnett2001tnfalpha,mason2001episodic}.

\section{Methods}
We categorized state-of-the-art segmentation methods not only on EM datasets, but also from challenges involving natural images. The state-of-the-art methods on EM dataset \cite{wei2020mitoem,lin2021nucmm} employ CNN to predict binary map\cite{ronneberger2015unet}, affinity map \cite{lee2017superhuman}, and extra targets include contours\cite{wei2020mitoem} and distance \cite{lin2021nucmm,wei2021axonem}. As shown in Fig. \ref{fig2}, CNN-based methods are unable to overcome certain challenges, including scale differences, Node of Ranvier, and long-range axons. Hence, we employed a vision transformer (ViT)-based model to address these limitations. We compared the publicly available pre-trained ViTs, including MAE \cite{he2021mae} and BEiT \cite{bao2021beit}. Additionally, we fine-tuned the image encoder at the base size of large-scale natural image segmentation model SAM \cite{kirillov2023segment} to EM-SAM. It is important to note that unlike \cite{ma2023segment,wu2023medical}, we removed prompt encoder and the neck part of image encoder from SAM's \cite{kirillov2023segment}, as our automatic segmentation pipeline does not involve any prompts. For more detailed information about the models, we refer the readers to the original papers.

\section{Experiments}
\subsection{Implementation Details}
To ensure a fair comparison of ViT-based models, we utilized the same decoder structure but incorporated 2D convolutional layers as proposed in \cite{hatamizadeh2022unetr}. During training, we applied the same data augmentation techniques and employed the Warmup Cosine Decay learning rate schedule \cite{loshchilov2016sgdr,goyal2017warmup}. We used the SGD optimizer and updated all the parameters, including the encoder. The total number of training iterations was set to 200,000 with a mini-batch size of 2. The input size for the pre-trained ViTs from MAE \cite{he2021mae} and BEiT \cite{bao2021beit} was 224x224, while for EM-SAM \cite{kirillov2023segment}, it was 1024x1024. We employed Binary Cross-Entropy (BCE) loss during training and conducted the training on 2 NVIDIA 3090 GPUs. As for the CNN-based method, we replicated the state-of-the-art multi-task learning-based U-Net \cite{lin2021nucmm,wei2020mitoem} in its 2D version. Moreover, we adopted mean Intersection over Union (mIoU) as the evaluation metric for qualitative analysis.

\subsection{Benchmark results on AxonCallosumEM dataset}
We assess the performance of the previous state-of-the-art approach (UNet \cite{lin2021nucmm}) and fine-tune the latest pre-training methods based on vit (MAE \cite{he2021mae}, BEiT \cite{bao2021beit}) on our AxonCallosumEM dataset. Additionally, we investigate the influence of random initialization (scratched ViT \cite{Dosovitskiy2021google_transformer_experiments}) and input size (EM-SAM \cite{kirillov2023segment}). Specifically, the AxonCallosumEM dataset is divided into training, validation, and testing subsets, consisting of the genu, the body, and the splenium of the corpus callosum. We determine the hyperparameters for each model using the validation dataset and evaluate their final performance on the testing dataset.

\begin{figure}
  \centering
  \includegraphics[width = 1\textwidth]{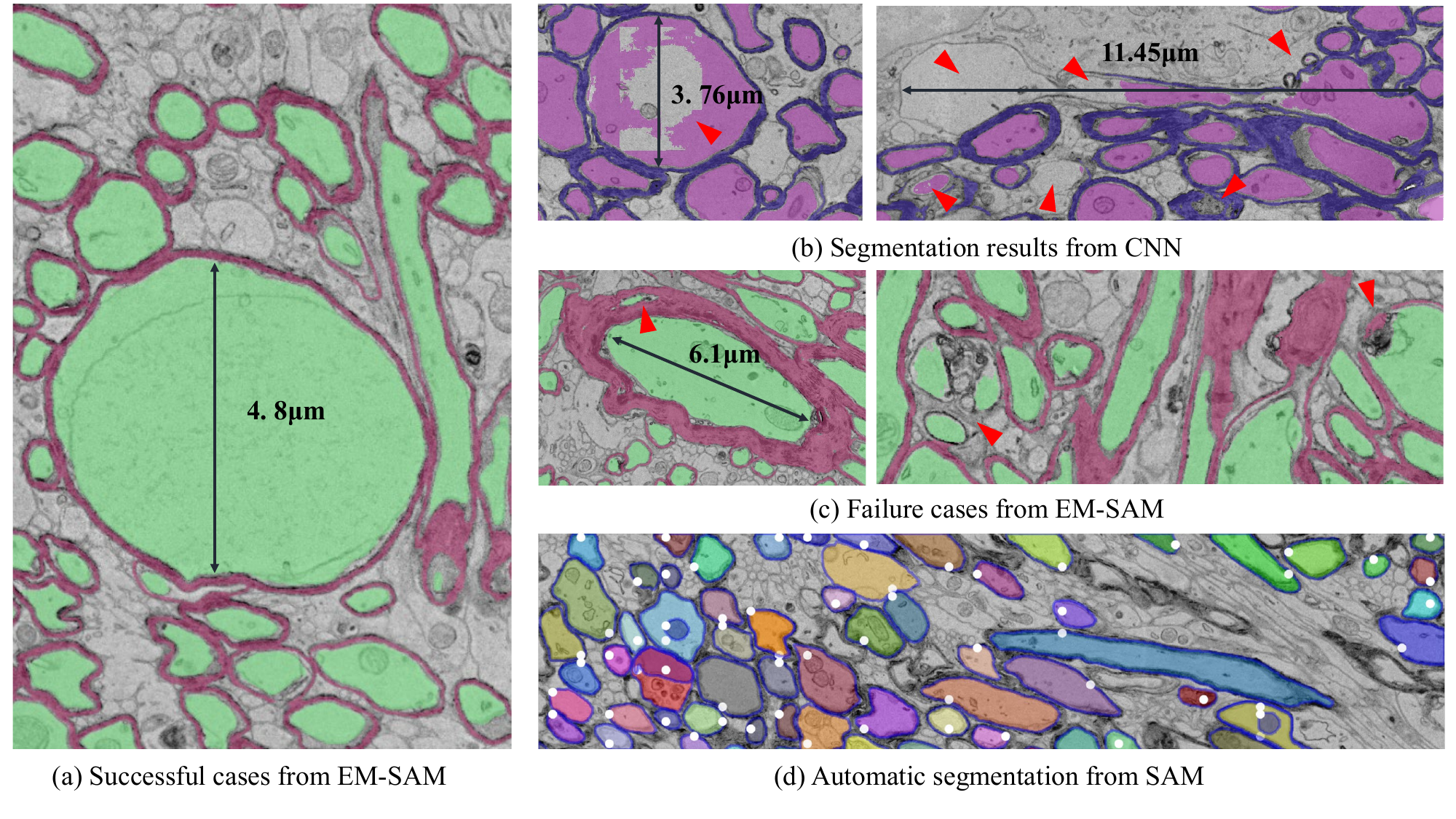}
  \caption{Qualitative outcomes of the AxonCallosumEM dataset are presented, showcasing the successful and failure cases of vit-based EM-SAM \cite{kirillov2023segment} in (a) and (c) respectively, while the results of CNN-based model \cite{lin2021nucmm} and original SAM is displayed in (b) and (d). The highlighted red arrows indicate segments with errors. (a) The demonstration of EM-SAM \cite{kirillov2023segment} model's ability to effectively address the segmentation challenges posed by large-scale and long-range objects, owing to its capacity to long-range dependencies. Conversely, in (b), the CNN-based model \cite{lin2021nucmm} fails to accurately segment objects of considerable size, such as the Node of Ranvier, as well as objects exhibiting complex morphology. Furthermore, (c) Red arrows demonstrate that since an incomplete understanding of the semantic information pertaining to myelinated axons and myelin sheath, the EM-SAM \cite{kirillov2023segment} model misclassifies gaps within the myelin sheath or the cell nucleus with thicker membranes as myelinated axons. (d) Without fine-tuning, the automatic segmentation mode of SAM \cite{kirillov2023segment} cannot categorize all the myelinated axons and myelin sheaths since the lack of understanding of related semantic information.}
  \label{fig4}
\end{figure}

\begin{table}
\caption{Benchmark results on AxonCallosumEM dataset. We compare state-of-the-art CNN-based and vit-based methods using mean Intersection over Union (mIoU).}
\centering
\begin{tabular}{cccc}
    \textbf{Method} & \textbf{Input size} & \textbf{mIoU} \\\toprule
    UNet \cite{lin2021nucmm} & $256\times256$ & 0.919  \\
    Scratched Vit-Base \cite{Dosovitskiy2021google_transformer_experiments} & $224\times224$ &  0.947  \\
    MAE-Base \cite{he2021mae}& $224\times224$ & 0.966 \\
    BEiT-Base \cite{bao2021beit}& $224\times224$ &  0.962\\
    EM-SAM-Base \cite{kirillov2023segment}& $1024\times1024$ & 0.984
    \\\bottomrule
    \label{table1} 
    \end{tabular}
\end{table}

As illustrated in Table \ref{table1}, the UNet-based method with multi-target prediction achieved the lowest mIoU. In Fig. \ref{fig4} (b), the UNet method demonstrates limitations in segmenting large or heterogeneous objects due to its limited receptive field. Nevertheless, it remains a reliable method as it accurately segments objects of common sizes. The ViT-based method \cite{Dosovitskiy2021google_transformer_experiments} exhibits a 0.028 improvement in mIoU compared to UNet \cite{lin2021nucmm}. Despite the smaller input size of the ViT-based method \cite{Dosovitskiy2021google_transformer_experiments} compared to UNet \cite{lin2021nucmm}, its ability to capture long-range dependencies through self-attention overcomes the challenge of scale differences among axons. Additionally, we explore the impact of fine-tuning publicly available pre-trained models and large-scale segmentation models trained on natural images. In Table \ref{table1}, fine-tuning the pre-trained MAE \cite{he2021mae} and BEiT \cite{bao2021beit} models yields enhancements of 0.019 and 0.015, respectively, compared to training from scratch. Furthermore, the fine-tuned EM-SAM \cite{kirillov2023segment} achieves state-of-the-art performance due to its original enormously large training dataset and $~4\times$ larger input size, comparing with MAE \cite{he2021mae} and BEiT \cite{bao2021beit}. As shown in Fig. \ref{fig4} (a), the fine-tuned EM-SAM \cite{kirillov2023segment} not only accurately segments large objects but also effectively handles challenges related to long-range morphology and demyelination. However, since the model does not fully comprehend the semantic information of myelinated axons and myelin sheaths, some segmentation errors persist. In Fig. \ref{fig4} (c), the EM-SAM \cite{kirillov2023segment} model misclassifies the gaps inside the myelin sheath or the cell nucleus with thicker membranes as myelinated axons, and struggles to identify boundaries and morphology with unclear features. Please noted that, only nature images were fed to the model of MAE \cite{he2021mae}, BEiT \cite{bao2021beit}, and EM-SAM \cite{kirillov2023segment} before our fine-tuning procedure. Moreover, our experimental results provide support for the adaption of large-scale natural image models to biomedical image modeling, enabling positive impacts for downstream tasks through fine-tuning (See Fig. \ref{fig4} (a) and (d)) \cite{raghu2019transfusion,frid2018improving,ma2023segment}.

\section{Conclusion}
In this paper, we introduce a large-scale dataset for semantic segmentation of myelinated axons and myelin sheaths, which reveals the limitations of current state-of-the-art methods in handling complex morphologies. Similar to ImageNet \cite{deng2009imagenet} for natural images, our densely annotated 2D AxonCallosumEM dataset holds potential for various applications beyond its original task, such as pre-training; adaptability across nature images modeling and biomedical images modeling; evaluating latest methods; and biomedical analyses across mouse models. Moreover, our EM-SAM \cite{kirillov2023segment} provides the idea of adapting large model of nature images processing to automatic high-throughput method of EM images processing.

\section{Conflict of Interest Statement}
The authors declare that the research was conducted in the absence of any commercial or financial relationships that could be construed as a potential conflict of interest.

\section{Acknowledgments}
We would like to thank the Prof. Qiu Zilong, from Shanghai Institutes for Biological Sciences, for providing the mouse of Rett Syndrome (RTT) model. 

\bibliographystyle{unsrt}  

\end{document}